\documentclass[11pt]{amsart}
\usepackage{amsmath}
\usepackage{amsxtra}
\usepackage{amscd}
\usepackage{amsthm}
\usepackage{amsfonts}
\usepackage{amssymb}
\usepackage{eucal}
\usepackage{epsfig}

\usepackage{latexsym,amsthm}

\DeclareFontEncoding{OT2}{}{}

\input epsf

\def\figin{\epsfcheck\figin}\def\figins{\epsfcheck\figins}
\def\epsfcheck{\ifx\epsfbox\UnDeFiNeD
\message{(NO epsf.tex, FIGURES WILL BE IGNORED)}
\gdef\figin##1{\vskip2in}\gdef\figins##1{\hskip.5in}
\else\message{(FIGURES WILL BE INCLUDED)}%
\gdef\figin##1{##1}\gdef\figins##1{##1}\fi}
\def\DefWarn#1{}
\def\figinsert{\goodbreak\topinsert}
\def\ifig#1#2#3#4{\DefWarn#1\xdef#1{fig.~\the\figno}
\writedef{#1\leftbracket fig.\noexpand~\the\figno}%
\figinsert\figin{\centerline{\epsfxsize=#3mm \epsfbox{#2}}}
\bigskip\medskip\centerline{\vbox{\baselineskip12pt
\advance\hsize by -1truein\noindent\footnotefont{\sl Fig.~\the\figno:}\sl\ #4}}
\bigskip\endinsert\noindent\global\advance\figno by1}

\def\half{\frac{1}{2}}

\theoremstyle{remark}

\theoremstyle{definition}

\numberwithin{equation}{section}

\newcommand{\Ref}[1]{{$($\ref{#1}$)$}}
\newcommand{\bean}{\begin{eqnarray}}
\newcommand{\eean}{\end{eqnarray}}
\newcommand{\be}{\begin{displaymath}}
\newcommand{\ee}{\end{displaymath}}
\newcommand{\bea}{\begin{eqnarray*}}
\newcommand{\eea}{\end{eqnarray*}}

\newcommand{\secref}[1]{Section~\ref{#1}}

\newcommand{\nc}{\newcommand}
\nc{\on}{\operatorname}
\nc{\p}{{\partial}}
\nc{\pa}{{\p}}
\nc{\ch}{\mbox{ch}}
\nc{\Z}{{\mathbb Z}}
\nc{\C}{{\mathbb C}}
\nc{\pone}{{\mathbb C\mathbb P}^1}

\nc{\CA}{{\mathcal A}}
\nc{\CB}{{\mathcal B}}
\nc{\CC}{{\mathcal C}}

\nc{\CE}{{\mathcal E}}
\nc{\CF}{{\mathcal F}}
\nc{\CG}{{\mathcal G}}
\nc{\CH}{{\mathcal H}}
\nc{\CI}{{\mathcal I}}
\nc{\CJ}{{\mathcal J}}
\nc{\CK}{{\mathcal K}}
\nc{\CL}{{\mathcal L}}
\nc{\CM}{{\mathcal M}}
\nc{\CN}{{\mathcal N}}
\nc{\CO}{{\mathcal O}}
\nc{\CP}{{\mathcal P}}
\nc{\CQ}{{\mathcal Q}}
\nc{\CR}{{\mathcal R}}
\nc{\CS}{{\mathcal S}}
\nc{\CT}{{\mathcal T}}
\nc{\CU}{{\mathcal U}}
\nc{\CV}{{\mathcal V}}
\nc{\CW}{{\mathcal W}}
\nc{\CX}{{\mathcal X}}
\nc{\CY}{{\mathcal Y}}
\nc{\CZ}{{\mathcal Z}}

\nc{\zb}{\ol{z}}
\nc{\xb}{{\bar x}}
\nc{\yb}{{\bar y}}
\nc{\ub}{{\bar u}}
\nc{\pb}{\bar\partial}
\nc{\wb}{\ol{w}}
\nc{\qb}{\ol{q}}
\nc{\nb}{\ol{n}}
\nc{\phb}{\ol\ph}
\nc{\jb}{\ol{j}}
\nc{\ib}{\ol{i}}
\nc{\kb}{\ol{k}}
\nc{\ph}{p}

\nc{\bi}{{\bf i}}
\nc{\bj}{{\bf j}}
\nc{\bk}{{\bf k}}
\nc{\bq}{{\bf q}}
\nc{\bv}{{\bf v}}

\nc{\dirac}{D\hspace*{-2.5mm}\slash}

\nc{\al}{\alpha}
\nc{\bt}{{\beta}}
\nc{\dl}{{\delta}}
\nc{\la}{\lambda}
\nc{\m}{\mu}
\nc{\ep}{\epsilon}
\nc{\si}{\sigma}
\nc{\om}{\omega}
\nc{\eps}{{\varepsilon}}
\nc{\bS}{{\mathbb S}}
\nc{\De}{\Delta}
\nc{\Ga}{\Gamma}
\nc{\La}{\Lambda}

\nc{\el}{\ell}

\nc{\arr}{\rightarrow}
\nc{\larr}{\longrightarrow}
\nc{\ri}{\rangle}
\nc{\lef}{\langle}
\nc{\su}{\widehat{{\mathfrak s}{\mathfrak l}}_2}
\nc{\sw}{{\mathfrak s}{\mathfrak l}}
\nc{\g}{{\mathfrak g}}
\nc{\h}{{\mathfrak h}}
\nc{\n}{{\mathfrak n}}
\nc{\N}{\widehat{\n}}
\nc{\G}{\widehat{\g}}
\nc{\gt}{\widetilde{\g}}
\nc{\one}{{\mathbf 1}}
\nc{\z}{{\mathfrak Z}}
\nc{\wt}{\widetilde}
\nc{\wh}{\widehat}
\nc{\cri}{_{\kappa_c}}
\nc{\kk}{h^\vee}
\nc{\sun}{\widehat{\sw}_N}
\nc{\ol}{\overline}
\nc{\ds}{\displaystyle}
\nc{\dzz}{\frac{dz}{z}}
\nc{\Res}{\on{Res}}
\nc{\mc}{\mathcal}
\nc{\Cal}{\mathcal}
\nc{\bb}{{\mathfrak b}}
\nc{\ot}{\otimes}
\nc{\R}{{\mathbb R}}
\nc{\yy}{{\mc Y}}
\nc{\ga}{\gamma}

\nc{\us}{\underset}
\nc{\opl}{\oplus}
\nc{\beq}{\begin{equation}}
\nc{\Rep}{\on{Rep}}
\nc{\sssec}{\subsubsection}
\nc{\ssec}{\subsection}
\nc{\lan}{\langle}
\nc{\ran}{\rangle}

\nc{\Vect}{\on{Vect}}
\nc{\ghat}{\G}
\nc{\T}{\mc T}
\nc{\Tloc}{\T^\g_{\on{loc}}}
\nc{\vac}{|0\ran}
\nc{\Wick}{{\mb :}}
\nc{\mb}{\mathbf}
\nc{\delz}{\partial_z}
\nc{\cali}{\mathcal}
\nc{\li}{\mathfrak l}
\nc{\lt}{\widetilde{\li}}
\nc{\astar}{a^*}
\nc{\cA}{{\mc A}}
\nc{\ka}{\kappa}

\nc{\OO}{{\mc O}}
\nc{\AutO}{\on{Aut} O}
\nc{\DerO}{\on{Der} O}
\nc{\DerpO}{\on{Der}_+ O}
\nc{\Au}{{\mc A}ut}
\nc{\mf}{\mathfrak}
\nc{\V}{{\mc V}}
\nc{\hh}{\wh{\h}}

\nc{\pp}{{\mathfrak p}}
\nc{\mm}{{\mathfrak m}}
\nc{\rr}{{\mathfrak r}}
\nc{\ket}{\rangle}
\nc{\zz}{{\mathfrak z}}
\nc{\gr}{\on{gr}}
\nc{\Spe}{\on{Spec}}
\nc{\rv}{\crho}
\nc{\can}{\on{can}}
\nc{\MOp}{\on{MOp}_G(D)}
\nc{\Db}{{\mathbb D}}
\nc{\ww}{w}

\nc{\Con}{\on{Conn}(\Omega^{\crho})_D}
\nc{\ConD}{\on{Conn}(\Omega^{\crho})_{\Db}}
\nc{\ConDL}{\on{Conn}(\Omega^{\rho})_{\Db}}
\nc{\ConDtL}{\on{Conn}(\Omega^{\rho})_{\Db^\times}}
\nc{\OpD}{\on{Op}_G(\Db)}
\nc{\crho}{\check{\rho}}
\nc{\chal}{\check{\al}}
\nc{\cchi}{\check{\chi}}
\nc{\cLa}{\check\Lambda}
\nc{\cla}{\check\la}
\nc{\cmu}{\check\mu}

\nc{\PP}{{\mathbb P}}
\nc{\TT}{{\mathbb T}}

\nc{\bone}{{\mb 1}}

\nc{\bs}{\backslash}

\def\tr{{\rm tr}}

\nc{\zzb}{z \zb}

\nc{\pf}{\int\hspace*{-3.5mm}\bs}

\nc{\inn}{\on{in}}
\nc{\out}{\on{out}}
\nc{\covac}{\langle 0|}
\nc{\ptwo}{{\mathbb C}{\mathbb P}^2}
\nc{\BS}{{\mathbb S}}

\begin{document}

\title{Notes on instantons in topological field theory and
beyond}\thanks{Supported by the DARPA grant HR0011-04-1-0031}

\author{E. Frenkel}

\address{Department of Mathematics, University of California,
       Berkeley, CA 94720, USA}

\author{A. Losev}

\address{Institute of Theoretical and Experimental Physics,
       B. Cheremushkinskaya 25, Moscow 117259, Russia}

\author{N. Nekrasov}

\address{Departments of Mathematics and Physics, Princeton University,
Princeton NJ 08544, USA and Institut des Hautes \'Etudes
Scientifiques, 35, Route de Chartres, Bures-sur-Yvette, F-91440,
France\footnote{permanent address. On leave of absence from: ITEP,
Moscow, Russia}}

\centerline{\hfill ITEP-TH-04/07 \ IHES-P/07/08}
\centerline{\hphantom{void}}
\centerline{\hphantom{void}}

\date{February 2007}

\begin{abstract}
This is a brief summary of our studies of quantum field theories in a
special limit in which the instantons are present, the anti-instantons
are absent, and the perturbative corrections are reduced to
one-loop. We analyze the corresponding models as full-fledged quantum
field theories, beyond their topological sector. We show that the
correlation functions of all, not only topological (or BPS),
observables may be studied explicitly in these models, and the
spectrum may be computed exactly. An interesting feature is that the
Hamiltonian is not always diagonalizable, but may have Jordan blocks,
which leads to the appearance of logarithms in the correlation
functions. We also find that in the models defined on K\"ahler
manifolds the space of states exhibits holomorphic factorization. In
particular, in dimensions two and four our theories are logarithmic
conformal field theories.

\end{abstract}

\maketitle

\section{Introduction}

Most two- and four-dimensional quantum field theories have two kinds
of coupling constants: the actual coupling $g$, which in particular
counts the loops in the perturbative calculations, and the topological
coupling, ${\vartheta}$, the theta-angle, which is the chemical
potential for the topological sectors in the path integral. These
couplings can be combined into the complex coupling ${\tau}$ and its
complex conjugate ${\tau}^{*}$. The idea is to study the dependence of
the theory on $\tau$, ${\tau}^{*}$ as if they were two separate
couplings, not necessarily complex conjugate to each other.

For example, in the four-dimensional gauge theory one combines the
Yang-Mills coupling $g$ and the theta-angle $\vartheta$ as follows:
\begin{equation}
{\tau} = \frac{\vartheta}{2\pi} + \frac{4{\pi}i}{g^{2}}.
\label{tauym}
\end{equation}
For the two dimensional sigma model with the complex target space $X$,
endowed with a Hermitian metric $g_{i\jb}$ and a $(1,1)$ type
two-form $B_{i\jb}$ one defines
\begin{equation}
{\tau}_{i\jb} = B_{i\jb} + i g_{i\jb}.
\label{tausigma}
\end{equation}
If $dB = 0$, then the two-form $B$ plays the role of the theta-angle.

A similar coupling constant may also be introduced in the quantum
mechanical model on a manifold $X$ endowed with a Morse function
$f$, with the Lagrangian
\begin{equation}
L = \frac{\la}{2} \left( g_{\m\nu} {\dot x}^{\m}{\dot x}^{\nu} +
g^{\m\nu} {\p}_{\m}f {\p}_{\nu} f \right) - i {\vartheta} {\p}_{\m} f
{\dot x}^{\m} + \ldots \, .
\label{qmlag}
\end{equation}
where $\ldots$ denote the possible fermionic terms (in the
supersymmetric version). The corresponding coupling $\tau$ may then be
taken to be
$$
\tau = {\vartheta} + i {\la}.
$$
For finite $\la$ the supersymmetric quantum mechanics with
the bosonic Lagrangian \Ref{qmlag} is the model studied by E.~Witten
in his proof of Morse inequalities \cite{W:morse}.

In all these examples we expect the correlators to be functions of
$\tau$, $\tau^{*}$. It is reasonable to expect that they may be
analytically continued to the domain of complex couplings $g$ and
$\vartheta$. In particular, the theory should greatly simplify in the
limit:
\begin{equation}
{\tau}^{*} \to - i \infty \ , \ \qquad {\tau} \ {\rm fixed}.
\label{ourlimit}
\end{equation}
This is the weak coupling limit, in which the theta-angle has a large
imaginary part.

The reason for this simplification is that the theory in this limit is
described by a first-order Lagrangian. The corresponding
path integral represents the ``delta-form'' supported on the instanton
moduli space, which is essentially {\em finite-dimensional}. More
precisely, the instanton moduli space has components labeled by the
appropriate ``instanton numbers'', and each component is
finite-dimensional (after dividing by the appropriate gauge symmetry
group). Therefore the correlation functions are expressed as linear
combinations of integrals over these finite-dimensional components of
the instanton moduli space.

When we move away from the special point $\tau^* = -i \infty$ (with
fixed $\tau$), both instantons and anti-instantons start contributing
to the correlation functions. The path integral becomes a
Mathai-Quillen representative of the Euler class of an appropriate
vector bundle over the instanton moduli space, which is ``smeared''
around the moduli space of instantons (like a Gaussian distribution),
see, e.g., \cite{Moore}. Therefore general correlation functions are
no longer represented by integrals over the finite-dimensional
instanton moduli spaces and become much more complicated.

In supersymmetric models there is an important class of observables,
called the {\em BPS observables}, whose correlation functions are
independent of $\tau^*$.  They commute with the supersymmetry charge
$Q$ of the theory and comprise the {\em topological sector} of the
theory. The perturbation away from the point $\tau^* = -i \infty$ is
given by a $Q$-exact operator, and therefore the correlation functions
of the BPS observables (which are $Q$-closed) remain unchanged when we
move away from the special point. This is the secret of success of the
computation of the correlation functions of the BPS observables
achieved in recent years in the framework of topological field theory:
the computation is actually done in the theory at $\tau^* = -i
\infty$, but because of the special properties of the BPS observables
the answer remains the same for other values of the coupling
constant. But for general observables the correlation functions do
change in a rather complicated way when we move away from the special
point.

We would like to go {\em beyond the topological sector} and consider
more general correlation functions of non-BPS observables. Why should
we be interested in these more general correlation functions? Here are
some of the reasons for doing this. Other reasons will become more
clear later.

\begin{itemize}

\item Understanding non-supersymmetric quantum field theories with
instantons: It is generally believed that realistic quantum field
theories should be viewed as non-supersymmetric phases of
supersymmetric ones. This means that the observables of the original
theory may be realized as observables of a supersymmetric theory. But
they are certainly not going to be BPS observables. Therefore we need
to develop methods for computing correlation functions of such
observables.

\item Elucidating the pure spinor approach to superstring theory:
Non-supersym\-metric versions of our models (such as the ``curved
$\beta\gamma$-systems'') play an important role in this approach
\cite{Berk}.

\item Constructing new invariants: The correlation functions in the
topological sector of the quantum field theories considered above give
rise to invariants of the underlying manifold, such as the
Gromov-Witten and Donaldson invariants. We hope that the correlation
functions of the full quantum field theory may allow us to detect some
finer information about its geometry.

\end{itemize}

Since our goal now is to understand the full quantum field theory, and
not just its topological sector, it is reasonable to try to describe
the theory first for special values of the coupling constants, where
the correlation functions are especially simple. It is natural to
start with the limit $\tau^* = -i \infty$ (with finite $\tau$). We may
then try to extend the results to a neighborhood of this special value
by perturbation theory. We hope that this will give us a viable
alternative to the conventional approach using the expansion around a
Gaussian point. The advantage of this alternative approach is that,
unlike in the Gaussian perturbation theory, we do not need to impose a
linear structure on the space of fields. On the contrary, the
non-linearity is preserved and is reflected in the moduli space of
instantons, over which we integrate in the limit $\tau^* \to -i
\infty$ (for more on this, see \secref{new obs}). This is why we
believe that our approach may be beneficial for understanding some of
the hard dynamical questions, such as confinement, that have proved to
be elusive in the conventional formalism.

The 4D $S$-{\em duality} (and its 2D analogue: the mirror symmetry)
gives us another tool for connecting our limit to the {\em physical
range of coupling constants}. In a physical theory, in which $\tau^*$
is complex conjugate to $\tau$, the $S$-duality sends $\tau \mapsto
-1/\tau$. It is reasonable to expect that $S$-duality still holds when
we complexify the coupling constants $\tau, \tau^*$. It should then
act as follows:
$$
\tau \mapsto -1/\tau, \qquad \tau^* \mapsto -1/\tau^*.
$$
Now observe that applying this transformation to $\tau^* = -i \infty$
and finite $\tau$, we obtain $\wt\tau = -1/\tau$ and $\wt\tau^* =
0$. These coupling constants are already within the range of physical
values, in the sense that both the coupling constant $g$ and the
theta-angle $\vartheta$ are finite! Therefore we hope that our
calculations in the theory with $\tau^* = -i \infty$ could be
linked by $S$-duality to exact non-perturbative results in a physical
theory beyond the topological sector.

In \cite{FLNi} (and the forthcoming companion papers \cite{FLNii}) we
have launched a program of systematic study of the $\tau^* \to -i
\infty$ limit of the instanton models in one, two and four
dimensions. In this paper we outline some of the salient features of
our constructions and results. We refer to the reader to the above
papers for more details and additional references.

\section{Lagrangian implementation of the limit $\tau^* \to -i
  \infty$}

In order to study the limit \Ref{ourlimit} properly we pass to the
first order formalism (after Wick rotating to Euclidean time and
completing the square). Then the Lagrangian \eqref{qmlag} becomes
\begin{equation}
L \to - i p_{\m} \left( {\dot x}^{\m} - v^{\m} \right) - i {\tau}
{\dot f} + \frac{1}{2\la} g^{\m\nu} p_{\m}p_{\nu},
\label{lagqmi}
\end{equation}
where $v^{\m}$ is the gradient vector field
\begin{equation}
v^{\m} = g^{\m\nu} {\p}_{\nu} f.
\label{grad}
\end{equation}
of the Morse function $f$.

Let us complexify $\vartheta$ and set
\begin{equation}    \label{tau qm}
\tau = {\vartheta} + i {\la}.
\end{equation}
As $\la \to \infty$ with $\tau$ fixed (so that $\tau^* \to -i \infty$)
the Lagrangian simplifies as follows:
\begin{equation}
L \to L_{\infty} = - i p_{\m} \left( {\dot x}^{\m} - v^{\m} \right) -
i {\tau} {\dot f}.
\label{laginf}
\end{equation}
Now, if we integrate the $p_{\m}$'s out, we immediately see that the
path integral localizes onto the union of finite-dimensional moduli
spaces ${\CM}_{a,b}$ of the gradient trajectories, i.e. solutions to
the equations:
\begin{equation}
\frac{dx^{\m}}{dt} = v^{\m}, \label{gradtr}
\end{equation}
obeying the boundary conditions
$$ 
x ( - \infty ) = a , \qquad x ( + \infty ) = b.
$$
Here $a,b$ are critical points of $f$, or, equivalently, the zeros
of the vector field $v$.

For the sake of simplicity, let us focus on a supersymmetric model.
Then in addition to the bosonic fields $x^{\m}(t), p_{\m}(t)$ we have
their fermionic partners ${\psi}^{\m}(t), {\pi}_{\m}(t)$. The
fermionic part of the Lagrangian is, roughly (we skip the couplings to
the connections on the tangent bundle etc.),
\begin{equation}
L_{\on{fermion}} = i {\pi}_{\m} \left( {\dot \psi}^{\m} -
{\p}_{\nu}v^{\m} {\psi}^{\nu} \right). \label{lagfer}
\end{equation}
In this case the determinants obtained by integrating over the
fluctuations around the solutions to \Ref{gradtr} cancel, leaving us
with the integral over the superspace ${\Pi}T{\CM}_{a,b}$, where the
odd directions come from the solutions to the fermion equations:
\begin{equation}
\frac{d{\psi}^{\m}}{dt} = {\p}_{\nu}v^{\m}(x(t)) {\psi}^{\nu}.
\label{fermod}
\end{equation}

\subsection{Lagrangian implementation: observables}

The general (local) observables correspond to functions 
$$
{\CO}(x, p, {\pi}, {\psi})
$$
which become differential operators on ${\Pi}TX$ upon quantization.
The simplest observables are the so-called {\em evaluation
observables}. They correspond to the functions ${\CO}(x, \psi)$ on
${\Pi}TX$, i.e., differential forms ${\varpi}$ on $X$. Their
correlation functions are the easiest to study:
\begin{equation}
\langle_{\kern -8pt a} \ {\CO}_{1}(t_{1}) \ldots {\CO}_{k} (t_{k})
\rangle_{b} = e^{-i {\tau}( f(a) - f(b))} \ \int_{{\CM}_{a,b}}
ev_{t_{1}}^{*}{\varpi}_{1} \wedge \ldots \wedge ev_{t_{k}}^{*}
{\varpi}_{k},
\label{correv}
\end{equation}
where 
$$
ev_{t}: {\CM}_{a,b} \to X
$$
is the evaluation of the gradient trajectory at the moment of time $t$:
\begin{equation}
ev_{t} \left[ x \right]  = x(t).
\label{eval}
\end{equation}
Note that for closed differential forms ${\varpi}_{i}$ such that
$d{\varpi}_{i}= 0$ the correlator \Ref{correv} is independent of the
time positions $t_i$ (at least in the case when $X$ is a K\"ahler
manifold) and defines a simplified version of the celebrated
Gromov-Witten invariants.

\section{Hamiltonian implementation of the ${\tau}^{*} \to -i\infty$
  limit}

For simplicity let us set temporarily $\tau = 0$ (see formula
\eqref{tau qm}). Therefore $\vartheta = -i\la$. The addition of the
topological term $-i \vartheta \int df = - \la \int df$ to the
Lagrangian (see formula \eqref{qmlag}) amounts to the following
redefinition of the wave-functions:
\begin{equation}
{\Psi} \mapsto {\Psi}^{\rm in} = {\Psi} e^{{\la}f}, \qquad {\Psi}
\mapsto {\Psi}^{\rm out} = {\Psi}^{*} e^{-{\la}f}.
\label{wvrdf}
\end{equation}
This maps the standard Hermitian inner product to the pairing
$$
{\langle} \Psi^{\rm out} | \Psi^{\rm in} \rangle = \int {\Psi}^{\rm
  out} \wedge {\Psi}^{\rm in}.
$$
Once $\vartheta$ is allowed to be complex, the manifest unitarity of the
usual quantum mechanics is lost, since ${\Psi}^{\rm out} \neq \left(
{\Psi}^{\rm in} \right)^{*}$. However, for finite $\la$ we can always
undo the transformation \Ref{wvrdf} and establish the isomorphism
between the spaces of in- and out-states.

The redefinition \eqref{wvrdf} of the wave-functions leads to the
following redefinition of observables:
$$
{\mathcal O} \mapsto {\mathcal O}^{\on{in}} = e^{\la f} {\mathcal O}
e^{-\la f}, \qquad {\mathcal O} \mapsto {\mathcal O}^{\on{out}} =
e^{-\la f} {\mathcal O} e^{\la f}.
$$
In particular, the Hamiltonian $ - \frac{1}{2\la} {\Delta} +
\frac{\la}{2} \Vert d f \Vert^{2}$ of the original quantum mechanical
model with the Lagrangian \Ref{qmlag} is mapped to
\begin{equation}
H^{\on{in}}_\la = {\CL}_{v} - \frac{1}{2\la} {\Delta}, \qquad
H^{\on{out}}_\la = - {\CL}_{v} - \frac{1}{2\la} {\Delta},
\label{newham}
\end{equation}
where ${\CL}_{v}$ is the Lie derivative along the gradient vector
field $v$.  In the limit ${\la} \to \infty$ they become
$H_{\infty}^{\on{in}} = {\CL}_{v}, H_{\infty}^{\on{out}} = -
{\CL}_{v}$.

More generally, if we wish to keep non-zero $\tau = {\vartheta} + i
{\la}$, then we consider the Hamiltonian
$$
{H}_{\tau,\tau^*} =   {1\over 2\la
} \{ d - i {\tau} df \wedge \ ,\ d^{*} + i {\tau}^{*} \iota_{v} \} 
$$
and then take the limit $\la \to +\infty, \vartheta \to -i \infty$,
while keeping $\tau$ fixed. Then the Hamiltonian tends to
\begin{equation}
H_{\tau} = {\CL}_{v} - i{\tau} \Vert v \Vert^2 = e^{i \tau f}
H_{\infty} e^{-i \tau f}.
\label{hamtau}
\end{equation}

\subsection{Local theory: harmonic oscillator}

Let us analyze the spectrum of the Hamiltonian \Ref{hamtau} near a
fixed point $x_0$, $v(x_{0}) =0$. The problem reduces to that of the
spectrum of harmonic oscillator. The only remaining issue is the
effect of the redefinition \Ref{wvrdf} on the well-known eigenstates
of the Hamiltonian $H = - \frac{1}{2\la} {\Delta} + \frac{\la}{2}
\Vert d f \Vert^{2}$

There are two basic cases, corresponding to a repulsive critical
point, with $f = {\om} x^{2}/2$, and an attractive critical point, with
$f = - {\om}x^{2}/2$, (we will assume that ${\om} > 0$). In the limit
${\la} \to \infty$ the states and the Hamiltonians are as follows. In
the repulsive case (we set $\tau =0$ for simplicity):
\begin{align}
\label{attract}
{\Psi}^{\rm in} &= P (x,dx), \qquad {\Psi}^{\rm out} = P (
{\p}_{x},\imath_{\pa_x} )
{\delta}(x) \\ \notag H^{\rm in} &= {\om} {\CL}_{x {\pa_x}},
\qquad H^{\rm out} = - {\om} {\CL}_{x {\pa_x}}.
\end{align}
In the attractive case:
\begin{align}
\label{repul}
{\Psi}^{\rm out} = P (x,dx), \qquad {\Psi}^{\rm in} = P (
 {\p}_{x},\imath_{\pa_x} )
 {\delta}(x) \\ \notag H^{\rm in} = -{\om} {\CL}_{x {\pa_x}},
 \qquad H^{\rm out} = + {\om} {\CL}_{x {\pa_x}},
\end{align}
where $P( \cdot )$ is a polynomial differential form. It is easy to
see that the spectrum of the Hamiltonian(s) is bounded from below. The
negative signs are compensated by the fact that the scaling dimensions
of the delta-function and its derivative are also negative. Thus in
all cases we get the eigenvalues-values:
\begin{equation}
E_{n} = | {\om} | n \ , \ n = 0, 1,2, \ldots \, .
\label{en}
\end{equation}
In the non-supersymmetric case there are corrections to the energy
levels of them form $\pm \half \omega$. But in all cases we obtain the
real spectrum bounded below.  More generally, we could have gotten
complex eigenvalues, but their real parts are always bounded below, as
is required by stability.

\subsection{Example of a global theory: two dimensional sphere.}

The next example illustrates general phenomenon of the state-mixing in
the presence of instantons. We study quantum mechanics on $X = {\mathbb
S}^{2}$. Take
\begin{equation}
f = {1\over 4} {{z\zb - 1} \over {z \zb +1 }}, \qquad g = {dz d\zb
\over ( 1+ z\zb )^{2}}.
\label{Morsefsph}
\end{equation}
The corresponding gradient vector field
\begin{equation}
v = z {\p}_{z} + {\zb}{\p}_{\zb}
\label{gradv}
\end{equation}
has two fixed points: $z=0$ and $z= \infty$.

We can cover $X$ with two coordinate patches, isomorphic to $\C$,
$$
\C_0 = \BS^2 \bs \{ \infty \}, \qquad \C_\infty = \BS^2 \bs \{ 0
\}.
$$
The coordinates $z$ and $w$ on these patches are related via $z
= 1/w$.  The moduli space of gradient trajectories splits as a
disjoint union of the following components:
\begin{equation}
{\CM}_{\infty, \infty} = \{ \infty \}, \qquad {\CM}_{\infty, 0} \simeq
{\C}^{\times}, \qquad {\CM}_{0,0} = \{ 0 \}.
\label{modsp}
\end{equation}
Quantum mechanically, we have a two-well potential with an additional
$U(1)$ symmetry. Let us denote the generator of the $U(1)$ rotations by
$P$,
\begin{equation}
P = - i \left( z {\p}_{z} - {\zb} {\p}_{\zb} \right).
\label{rot}
\end{equation}
We can lift the degeneracy caused by the $U(1)$ symmetry by studying
the common spectrum of the operators $( L_{0} , {\bar L}_{0} ) =
{\half} \left( H + i P , H - i P \right)$.

What about the two wells? Naively, in the $\la \to \infty$ limit the
spectrum should be well approximated by the harmonic oscillators
corresponding to the two minima of the potential. The function $f$
behaves as
$$
f \sim - {1\over 4} + {\half} z{\zb} \qquad \on{near} \qquad z=0
$$
(repulsive critical point), and as
$$
f \sim {1\over 4} - {\half} w{\wb} \qquad \on{near} \qquad z = \infty
$$
(attractive critical point). Therefore the analysis of the previous
section leads us to the following description of the spaces of ``in''
states:
\begin{align} \label{hsph}
\qquad {\mc H}^{\rm in} &= {\mc H}^{\rm in}_{\C_0} \oplus {\mc H}^{\rm
in}_{\infty}, \\ \notag {\mc H}^{\rm in}_{\C_0} &= {\C} [ z, {\zb} ,
dz, d{\zb} ], \\ \notag {\mc H}^{\rm in}_{\infty} &= {\C} [ {\p}_{w},
{\p}_{\wb} , \iota_{{\p}_{w}} , \iota_{{\p}_{\wb}} ] {\delta}^{(2)}(w,
{\wb}) dw \wedge d{\wb},
\end{align}
and ``out'' states:
\begin{align} \label{hsph1}
{\mc H}^{\rm out} &= {\mc H}^{\rm
out}_{0} \oplus {\mc H}^{\rm out}_{\infty}\hfill, \\ \notag {\mc
H}^{\rm out}_{\infty} &= {\C} [ w, {\wb} , dw, d{\wb} ], \\ \notag
{\mc H}^{\rm out}_{0} &= {\C} [ {\p}_{z}, {\p}_{\zb} ,
\iota_{{\p}_{z}} , \iota_{{\p}_{\zb}} ] {\delta}^{(2)}(z, {\zb}) dz
\wedge d{\zb}.
\end{align}
This gives us the following energy levels:
\begin{align}    \label{spec}
{\rm on} \ {\mc H}^{\rm in}_{\C_0} \ : \quad
n + \ol{n} &\qquad {\rm on} \qquad z^{n}{\zb}^{\ol{n}} \\ \notag
n + \ol{n}+1 & \qquad {\rm on} \qquad z^{n}{\zb}^{\ol{n}} dz \;
\on{and} \; z^{n}{\zb}^{\ol{n}}
d{\zb} \\ \notag n + \ol{n}+2 & \qquad {\rm on} \qquad
z^{n}{\zb}^{\ol{n}} dz\wedge d{\zb} \\ \notag
\qquad {\rm on} \ {\mc H}^{\rm in}_{\infty} \ : \quad
n + \ol{n} +2 & \qquad  {\rm on} \qquad
{\pa}_{w}^{n}{\pa}_{\wb}^{\ol{n}} {\delta}^{(2)}(w,{\wb} )\\ \notag
n + \ol{n}+1 & \qquad {\rm on} \qquad
{\pa}_{w}^{n}{\pa}_{\wb}^{\ol{n}} {\delta}^{(2)}(w,{\wb} )dw \;
\on{and} \;
{\pa}_{w}^{n}{\pa}_{\wb}^{\ol{n}} {\delta}^{(2)}(w,{\wb}) d{\wb}
 \\ \notag 
n + \ol{n} & \qquad {\rm on} \qquad
{\pa}_{w}^{n}{\pa}_{\wb}^{\ol{n}} {\delta}^{(2)}(w,{\wb}) dw\wedge
d{\wb}.
\end{align}
We have a similar description of ${\mc H}^{\rm out}$.
 
\subsection{Global theory: problems and their resolution}

Our description \Ref{hsph}--\Ref{spec} of the spaces
of states raises some serious questions. First of all, the
eigenfunctions $z^{n} {\zb}^{\ol{n}}$ are not well-defined on the
sphere, only on the big cell ${\C}_{0}$. Secondly, we have a
degenerate spectrum (except for the ground states in the zero- and
two-form sectors), and this contradicts the usual expectation that the
instantons lift the degeneracy of the spectrum.

It turns out that these problems can be resolved. To begin with, note
that we have the limiting wave-functions represented by the
delta-functions and their derivatives, which are not functions on the
sphere, either. Since we believe that these delta-functions are honest
limits of the wave-functions when $\la \to \infty$, then we have to
allow generalized functions (or distributions) as legitimate
wave-functions in this limit. This gives us a hint that we should try
to view those wave-functions that are polynomials in $z$ as generalized
functions as well.  This is possible, but there is a subtlety, which
leads to the appearance of Jordan blocks in the Hamiltonian.

So we wish to think of a polynomial $P(z, {\zb})$ as a distribution on
the smooth differential two-forms on ${\mathbb S}^{2}$. Of course, the
pole at infinity makes the naive integral of the product of $P$ and a
smooth two-form $\omega$ ill-defined (unless $\omega$ rapidly decays
at $\infty$). But let us regularize this integral by setting
\begin{equation}
\label{epst}
\langle z^{n}{\zb}^{\nb}, \omega \rangle = \left( \int_{|z|<
  {\epsilon}^{-1}} \ z^{n}{\zb}^{\nb}\  {\omega}
\right)_{{\epsilon}^{0}}.
\end{equation}
The integral on the right hand side may be written as
\begin{equation}    \label{C log}
C_0 + \sum_{i>0} C_i \ep^{-i} + C_{\log} \log \ep + o(1),
\end{equation}
where the $C_i$'s and $C_{\log}$ are some numbers. The right hand side
of \eqref{epst} is, by definition, the Hadamard {\em partie finie} of
the above integral, i.e., the constant coefficient $C_0$ obtained
after discarding the terms with negative powers of $\ep$ and $\log
\ep$ in the integral \eqref{epst} and taking the limit $\ep \to 0$
(this is also reminiscent of the Epstein-Glaser regularization
familiar in quantum field theory).

Note that this pairing is not canonical. Should we change $\epsilon$
to $2\epsilon$, the result will change by
$$
{\rm log}(2) \ \frac{1}{(n-1)!} \frac{1}{(\ol{n}-1)!}
{\pa}_{w}^{n-1} {\pa}_{\wb}^{\ol{n}-1} \left( \frac{{\omega}}{dwd{\wb}}
\right)\vert_{w={\wb} =0}, \qquad n,\ol{n}>0.
$$
Thus, we cannot separate the monomial $z^{n}{\zb}^{\nb}$, considered
as a distribution in the above sense, from the delta-like distribution
$\frac{1}{(n-1)!{(\nb-1)}!} {\pa}_{w}^{n-1}
{\pa}_{\wb}^{\nb-1} {\delta}^{(2)}(w,{\wb})$ (unless $n=0$ or
$\nb=0$). Thus, we observe a "mixing" between the states
$$
z^{n}{\zb}^{\nb} \qquad \on{and} \qquad \frac{1}{(n-1)!{(\nb-1)}!}
{\pa}_{w}^{n-1} {\pa}_{\wb}^{\nb-1} {\delta}^{(2)}(w,{\wb}).
$$
This is an instanton effect, as one can see clearly from the following
calculation.

Let us calculate the correlation function of the following
evaluation observables: a function $h$ and a two-form $\psi$ (it is
important in this calculation that $h$ is not closed, i.e., is not a
BPS observable). We have
\begin{equation}
\langle_{\kern -14pt \infty } \ {\CO}_{h}(0)
{\CO}_{\psi}(t) \rangle_{0} =
e^{i {\tau} \over 2} \int_{{\C}^{\times}} h\left( ze^{-t}
,{\zb}e^{-t}\right) {\psi} (z,{\zb}).
\label{corrl}
\end{equation}
(the $e^{i{\tau} \over 2}$-factor comes from the instanton part of the
action, $-i\tau \int_0^\infty df$). The energy spectrum can be
extracted by studying the $t$-dependence of the correlator
\Ref{corrl}. Take, for example, the following function and
two-form:
\begin{equation}
h = {1\over 1 + | z|^{2}}, \qquad {\psi} = \frac{dz \wedge d{\bar
z}}{\left( 1 + |z|^{2} \right)^{2}}.
\label{fpsi}
\end{equation}
 The correlator \Ref{corrl} equals
\begin{equation}
 \langle_{\kern -14pt \infty } \ {\CO}_{h}(0) {\CO}_{\psi}(t)
 \rangle_{0} = e^{i{\tau} \over 2} \left( - {1\over 1 - e^{-2t}} + {2
 t \over ( 1 - e^{-2t} )^{2} } \right).
 \label{corrlt}
 \end{equation}
Naively one would expect the $t$-dependence of the form:
\begin{equation}
\langle_{\kern -14pt \infty } \ {\CO}_{h}(0) {\CO}_{\psi}(t) \rangle_{0} 
= \sum_{\al} e^{ - t E_{\al}} \ h_{0 , \al} {\psi}_{\infty, \al}.
\label{tde}
\end{equation}
where $h_{0, \al}$ is the form-factor, the matrix element of the
operator $h$ between the vacuum associated to the point $0$, and the
eigenstate of the Hamiltonian with the energy level $E_{\al}$, and
$\psi_{\infty, \al}$ is the form-factor of $\psi$ between this
eigenstate and the covacuum associated to the point $\infty$.

The presence of the $t$-factor in \Ref{tde} implies that the
Hamiltonian is not diagonalizable! Instead, it has Jordan blocks:
\begin{equation}
{\rm exp} \left( \ t \cdot \begin{pmatrix} E & e^{i{\tau} \over 2} \\
0 & E \end{pmatrix} \ \right) =
\begin{pmatrix} e^{t E } & t e^{i{\tau} \over 2} e^{t E} \\ 0 & e^{t
    E} \end{pmatrix}.
\label{evol}
\end{equation}  
The reason why we got a Jordan block, as opposed to a matrix with a
non-zero entry under the diagonal (which would be diagonalizable with
slightly different eigenvalues), is the absence of anti-instantons in
our model. If they were present, the anti-instantons would contribute
a small matrix element under the diagonal in the Hamiltonian and hence
in the evolution operator.

A closer inspection shows that the Hamiltonian $H = z {\pa}_{z} +
{\zb} {\pa}_{\zb}$ acts on the states as follows:
\begin{align}
H \cdot {\pa}_{w}^{n}{\pa}_{\wb}^{\ol{n}}
{\delta}^{(2)}(w,{\wb} ) = ( n + \ol{n} +2 )
{\pa}_{w}^{n}{\pa}_{\wb}^{\ol{n}} {\delta}^{(2)}(w,{\wb} ), \\ H
\cdot z^{n+1} {\zb}^{\ol{n}+1} = ( n + \ol{n} +2 )
\left( z^{n+1} {\zb}^{\ol{n}+1} \right) + 
{\pa}_{w}^{n}{\pa}_{\wb}^{\ol{n}} {\delta}^{(2)}(w,{\wb} ).
\label{ham}
\end{align}
The mechanism generating the shift in the last line on \Ref{ham} is
the presence of the ${\rm log}{\epsilon}$ terms in the regularized
integrals \Ref{C log}. Here we assume that $\tau=0$. For general
$\tau$ there is a factor $e^{i {\tau}\over 2}$ in \Ref{evol}. For
simply-connected $X$ this factor can always be removed by changing the
basis in our space of states. Therefore, ${\tau}$ is not an observable
quantity for simply-connected $X$. In the case of non-simply connected
targets (and, in particular, in the 2D sigma models and 4D gauge
theories discussed below), the $\tau$-dependence of the Hamiltonian is
physical and observable.

\subsection{General K\"ahler target manifolds}    \label{gen kahler}

The above analysis generalizes in a straightforward way to the
supersymmetric quantum mechanical models on a K\"ahler manifold $X$
equipped with a holomorphic vector field $\xi$ coming from a
holomorphic torus action on $X$ with a non-empty set of isolated fixed
points (see \cite{FLNi}). Under our
assumptions, there is a {\em Bialynicki-Birula decomposition}
\cite{BB}
$$
X = \bigsqcup_{\al \in A} X_\al
$$
of $X$ into complex submanifolds $X_\al$, isomorphic to $\C^{n_\al}$,
defined as follows:
$$
X_\al = \{ x \in X | \lim_{t \to 0} \phi(t) \cdot x = x_\al \},
$$
where $\phi$ is the one-parameter subgroup corresponding to $\xi$.
We have the spaces ${\mc H}^{\on{in}}$ and ${\mc H}^{\on{out}}$ of
``in'' and ``out'' states, respectively. The former decomposes as a
direct sum
\begin{equation}
{\mc H}^{\on{in}} = \bigoplus_{\al \in A} {\mc H}^{\on{in}}_\al,
\end{equation}
where ${\CH}_\al$ be the space of {\em delta-forms supported on}
$X_\al$. An example of such a delta-form is the distribution on the
space of differential forms on $X$ which is defined by the following
formula:
\begin{equation}    \label{Delal}
\langle \Delta_\al,\eta \rangle = \int_{X_\al} \eta|_{X_\al}, \qquad
\eta \in \Omega^\bullet(X).
\end{equation}
All other delta-forms supported on $X_\al$ may be obtained by
applying to $\Delta_\al$ differential operators defined on a small
neighborhood of $X_\al$. The space ${\CH}^{\on{in}}_\al$ is graded by
the degree of the differential form. We have a similar description of
${\mc H}^{\on{out}}$.

The Hamiltonian is equal to ${\mc L}_\xi + {\mc L}_{\ol\xi}$ plus the
sum of off-diagonal terms which may be expressed in terms of
Grothendieck--Cousin operators corresponding to adjacent cells in the
above decomposition of $X$. These terms give rise to Jordan blocks in
the hamiltonians, as in the case of $\BS^2 = \pone$ analyzed above.

The correlation functions of these models may be computed both in the
Lagrangian approach (as integrals over moduli spaces of instantons)
and in the Hamiltonian approach (as matrix elements of operators
acting on the space of states). The factorization of the correlation
functions over intermediate states then leads to some interesting and
non-trivial identities on distributions, which are discussed in
detail in \cite{FLNi}.

\section{Infinite-dimensional versions: two and four dimensions}

The two-dimensional sigma model with the target manifold $X$ on the
worldsheets of genus zero and one may be analyzed along the same lines
as above by interpreting it as a quantum mechanical model on (a
covering of) the loop space $LX$. Similarly, the four-dimensional
gauge theory on ${\R}^{4}$, ${\BS}^{3} \times {\BS}^{1}$, $\ldots$,
may be interpreted as quantum mechanics on the space of gauge
equivalence classes on the three dimensional sphere. In this section
we discuss this briefly. Details will appear in \cite{FLNii}.

\subsection{Novikov, Morse-Bott, equivariant Morse...}

These theories have important subtleties. The corresponding functions
$f$ are multi-valued:
\begin{align*}
f &= \int_{{\BS}^{1}} d^{-1} {\omega} \qquad \text{in sigma models}, \\
f &= \int_{{\BS}^{3}} {\tr} \left( A dA + {2\over 3} A^{3} \right)
\qquad \text{in YM theory},
\end{align*}
so they are really Morse-Novikov functions. They may have non-isolated
critical points, like the constant loops in the sigma models, so they
are in fact Morse-Bott-Novikov functions. In addition, the above
Chern-Simons functional should be viewed as a Morse function in the
equivariant sense (due to gauge symmetry of connections). Because of
this, some adjustments need to be made in the formalism discussed
above. We will not discuss these models in detail here, referring the
reader to \cite{FLNi,FLNii}. We will only give some sample
calculations of the correlation functions which show that these models
also exhibit logarithmic behavior. This means that the Hamiltonian has
Jordan block. For simplicity we consider below twisted models, the
twisted ${\mc N} = (2,2)$ sigma models and twisted ${\mathcal N}=2$
gauge theory in four dimensions, also known as the Donaldson--Witten
theory. Again, we stress that we study these models as {\it
full-fledged} supersymmetric quantum field theories, not merely as
{\it topological field theories}.

\subsection{Sigma models.}

\subsubsection{Infinite radius limit.}

The twisted two dimensional sigma model with complex target space $X$
is described with the help of the following fields: $X^{\m} = (x^{i},
{\ol{x}}^{\jb}) : {\Sigma} \to X$, the momenta $p_{iw}, p_{\jb\wb}$,
the scalar fermions ${\psi}^{i}, {\bar\psi}^{\jb}$, and their momenta
${\pi}_{iw}, {\bar\pi}_{\jb\wb}$. The Lagrangian, written in the first
order form, reads:
\begin{align}    \label{lagr}
L &= - i \left( p_{iw} {\pa}_{\wb} x^{i} + p_{\jb\wb} {\pa}_{w}
{\ol{x}}^{\jb} + {\pi}_{iw} {\pa}_{\wb} {\psi}^{i} + {\bar\pi}_{\jb\wb}
{\pa}_{w} {\bar\psi}^{\jb} \right) \\ \notag &+ h^{i\jb}p_{iw}
p_{\jb\wb} + {\half} g_{i\jb} \left( {\pa}_w x^{i} {\pa}_{\wb}
{\ol{x}}^{\jb} - {\pa}_{\wb} x^{i} {\pa}_{w} {\ol{x}}^{\jb} \right) \\
\notag &+ \ {i\over 2} \ B_{\m\nu} \left( {\pa}_w X^{\m}
{\pa}_{\wb} X^{\nu} - {\pa}_{\wb} X^{\m} {\pa}_{w} X^{\nu} \right) + \
{\rm fermions}.
\end{align}
Upon elimination of the momenta $p_{iw}, p_{\jb\wb}$ the action
\Ref{lagr} turns into the standard Lagrangian of the type A twisted
supersymmetric sigma model with the target space $X$ endowed with the
Hermitian metric $g_{i\jb}$ and the $B$-field $B_{\mu\nu}$ (see
\cite{W:tsm}).

If the $B$-field is shifted by the imaginary $B$-field 
\begin{equation}
B \longrightarrow {\tau} = B + g_{i\jb} dx^{i} \wedge d{\ol{x}}^{\jb}
\label{btau}
\end{equation}
and the inverse metric $g^{i\jb}$ is sent to zero with $\tau$ kept
finite (so that $\tau^* \to \infty$), we obtain the Lagrangian
\Ref{lagr} of the {\it curved} $\beta\gamma$-$bc$ system, up to the
term $\int \ X^{*}{\tau}$. This is a version of the ``infinite radius
limit'' of this sigma model.

Now let us consider the case where $d{\tau} = 0$ (for $X$ is K\"ahler
this is what we get starting with the model \Ref{lagr}, for closed
$B$, $dB=0$). In this case the term $\int \ X^{*}{\tau}$ is
topological.

Suppose that $X$ is covered by coordinate patches $X = \cup_{\al}
{\mathcal U}_{\al}$. The model with the Lagrangian
\begin{equation}
L = - i \left( p_{iw} {\pa}_{\wb} x^{i} + p_{\jb\wb} {\pa}_{w}
{\ol{x}}^{\jb} + {\pi}_{iw} {\pa}_{\wb} {\psi}^{i} + {\bar\pi}_{\jb\wb}
{\pa}_{w} {\bar\psi}^{\jb} \right),
\label{bgbc}
\end{equation}
restricted to the maps which land in ${\mathcal U}_{\al}$, for some
$\al$, is the free $\beta\gamma$-$bc$ system, a $c=0$ superconformal
field theory. Thus, we can relate the chiral algebra of the sigma
model in the infinite radius limit to the {\it chiral de Rham complex}
\cite{MSV}.

\subsubsection{Instanton corrections.}

In contrast to most of the mathematical literature, we are not
interested in this chiral algebra {\it per se}. Rather, we are
interested in the full quantum field theory in the infinite radius
limit $\tau^* \to \infty$, in which the chiral and anti-chiral
sectors are combined in a non-trivial way.

We claim that just like in the quantum mechanical model, this global
theory, i.e., the sigma model with the instanton corrections, has a
non-diagonalizable Hamiltonian. As in the quantum mechanical case, the
spectrum of the Hamiltonian can be read off the correlation functions.
In the case of the sigma model the Hamiltonian is $L_{0} + {\bar
L}_{0}$, the sum of the chiral and anti-chiral Virasoro
generators. The Jordan block nature of the Hamiltonian implies that
the sigma model \Ref{bgbc} is a logarithmic conformal field theory
(LCFT). Note that the logarithmic corrections to the Virasoro
generators have non-perturbative character: they are caused directly
by the instantons!

However, we stress that the Hamiltonian is diagonalizable (in fact, is
identically equal to zero) on the BPS states. Therefore correlation
functions of the BPS observables which have been extensively studied
in the literature (and which are closely related to the Gromov-Witten
invariants) do not contain logarithms. In order to observe the
appearance of logarithms, we must consider non-BPS observables. The
Hamiltonian is also diagonalizable on all purely chiral (and
anti-chiral) states; thus, the chiral algebra of the theory is free of
logarithms.

Let us consider as an example the target manifold $X = {\C\mathbb
P}^{1}$. The instantons are labeled by a non-negative integer in this
case. The moduli space of degree $d$ instantons ${\mathcal M}_{d}$ has
complex dimension $2d+1$. We consider $d=1$.  Then the corresponding
moduli space ${\mathcal M}_{1} \simeq PGL_{2}({\C})$.  Consider the
correlator of the following evaluation observables:
\begin{equation}
\langle {\CO}_{{\omega}_{0}}(0) {\CO}_{{\omega}_{\infty}}({\infty})
	{\CO}_{{\omega}_{\rm FS}}(1) {\CO}_{h}(e^{-t})
	\rangle_{d=1},
\label{corrsm}
\end{equation}
where
$$
{\omega}_{0} = {\delta}^{(2)}(x) d^{2}x \ , \quad
{\omega}_{\infty} = {\delta}^{(2)}\left({1\over x}\right)
\frac{d^{2}x}{\vert x \vert^{4}} \ , \quad
{\omega}_{\rm FS} = \frac{d^{2}x}{( 1+ \vert x \vert^{2})^{2}},
$$
\begin{equation}
h = \frac{1}{1+ \vert x \vert^{2}}.
\label{omf}
\end{equation}
The delta-function two-forms ${\om}_{0}$, ${\om}_{\infty}$, supported
at $x=0$ and $x=\infty$, respectively, reduce the integration over
${\mathcal M}_{1}$ to that over the locus consisting of the maps:
\begin{equation}
x(w) = A w \ .
\label{redm}
\end{equation}

Thus, \Ref{corrsm} is equal to:
\begin{equation}
q \int \ \frac{d^{2}A}{( 1 + | A |^{2})^{2}} \frac{1}{1+ e^{-2t}
|A|^{2}} \propto q \left( \frac{-1}{1 - e^{-2t}} + \frac{2 t
e^{-2t}}{(1-e^{-2t})^{2}} \right),
\label{corrt}
\end{equation}
where $q$ is the instanton factor.  The $t$-dependence in \Ref{corrt}
implies the logarithmic nature of the two dimensional conformal
theory, in the same way as in the case of the quantum mechanical
models analyzed above. We conclude that the twisted sigma model on
${\C\mathbb P}^{1}$ is a logarithmic conformal field theory.

The space of states of the sigma model may be described in terms of
delta-forms supported on the ``semi-infinite'' strata of a
decomposition of the universal cover of $L\pone$, similarly to the
quantum mechanical models (see \secref{gen kahler}). 
We have a similar description of the sigma models associated to other
K\"ahler manifolds.

\subsection{Logarithmic conformal field theory in four dimensions.}

Four-dimensional ${\mathcal N}=2$ gauge theory with a compact gauge
group $G$ can be twisted just like the ${\mc N}=(2,2)$ two-dimensional
sigma model. The fields of the twisted theory are as follows. The
bosons: the gauge field $A_{m}$, the complex Higgs field ${\phi}$,
${\bar\phi} = {\phi}^{*}$, in the adjoint representation; and the
fermions, all in the adjoint representation: the one-form
${\psi}_{m}$, the self-dual two-form ${\chi}_{mn}^{+}$, the scalar
${\eta}$. By analogy with the quantum mechanics and the
two-dimensional sigma models we introduce a momentum $p^{+}_{mn}$ --
the self-dual two-form valued in the adjoint representation.

The super-Yang-Mills theory can be studied in the limit $g_{\rm YM}
\to 0$, with
$$
{\tau} = {\vartheta \over 2\pi} + {4\pi i \over g_{\rm YM}^{2}}
$$
kept finite (so that $\tau^* \to -i \infty$). The action of the
limiting theory looks as follows:
\begin{multline}
\label{sdsym}
S_{\rm ssdYM} = \\ \int\ {\tr} \left( -i p^{+} \wedge F + i {\chi}^{+}
\wedge D_{A}^{+} {\psi} + {\eta} \wedge \star D^{*}_{A} {\psi} + D_{A}
{\phi} \wedge \star D_{A} {\bar\phi} + [ {\psi} , \star {\psi} ]
{\bar\phi} \right) -\frac{i {\tau}}{4\pi} \int\ {\tr} F \wedge F.
\end{multline}
The action \Ref{sdsym} makes sense on any manifold ${\bf M}^{4}$. On
any ${\bf M}^{4}$ this theory has at least one fermionic symmetry,
generated by a scalar supercharge $\CQ$ \cite{W:tft}.  The usual
feature of such a theory is the $\CQ$-exactness of the stress-energy
tensor, which follows from the fact that all the metric dependence of
\Ref{sdsym} is contained in the $\CQ$-exact terms in the Lagrangian.

The conformal invariance of \Ref{sdsym} is much less appreciated in
the physics literature. Let us assume that ${\phi}$ is a
scalar, degree zero field, while $\ol{\phi}$ and $\eta$ are
half-densities, i.e. transform like ${\rm vol}_{\rm g}^{\frac{1}{2}}$,
under the coordinate transformations.  Then \Ref{sdsym} can be
rewritten, with explicit metric dependence, as:
\begin{multline}
\label{sdsymcft}
S_{\rm ssdYM} = {\CQ} \int\ -i g^{mm'} g^{nn'} g^{\half} {\tr}
\left( {\chi}^{+}_{mn} F_{m'n'} \right) +
g^{mn}g^{\frac{1}{4}} {\tr} \left( {\psi}_{n} D_{m} \ol{\phi}
\right) \\ -\frac{1}{4} {\CQ} \int\ g^{mn}g^{\frac{1}{4}} {\tr}
\left( {\psi}_{m} {\ol{\phi}} \right) {\pa}_{n} {\rm log} g
-\frac{i {\tau}}{4\pi} \int_{{\bf M}^{4}} {\tr} F \wedge F,
\end{multline}
where $D_{m} = {\pa}_{m} + [ A_{m}, \cdot ]$. The first line in
\Ref{sdsymcft} already defines a nice measure on the space of
fields. To make the theory explicitly conformally invariant 
we modify the stress-energy
tensor in a way analogous to the background charge modification of
the bosonic free field stress tensor in two dimensions $T \to T  + {\pa} J$, 
where $J_{m} \sim {\CQ} {\tr} \left( {\psi}_{m} \bar\phi \right)$.

{}The path integral in the theory \Ref{sdsym} localizes onto the
instanton moduli space of anti-self dual gauge field configurations,
i.e., the solutions of the equation
\begin{equation}
F_{A}^{+} = 0.
\label{asd}
\end{equation}
We now wish to apply our techniques to this gauge theory and
demonstrate its logarithmic nature (when we look beyond the
topological sector).

Consider the correlation function
\begin{equation}
C ( x, y ; z) = \langle {\CO}(x) {\CO}(y) {\mathcal S}(z) \rangle,
\end{equation}
where
$$
{\CO}(x) = {\tr}{\phi}^{2} \  , \qquad {\mathcal S}(z) = {\tr}
F_{mn}F^{mn}.
$$
We find that
$$
C(x,y ; z) = {1\over | x - y |^{4}} {\mathcal C} \left( {\left( {\xb}
  - {\yb}\right) \cdot \left( x + y - 2z \right) \over | x - y |^{2} }
\right),
$$
where we use the quaternionic notations, $x, y, z \in {\mathbb H}$,
and
\begin{equation}    \label{corrq}
{\mathcal C}(q) \propto
\frac{1}{|1-q|^{6}} \int_{0}^{1}
\frac{du}{M^{8}} 
\left( M P_{4}(M) - 3 (  1 + M)^2 (7 + 6 M + M^2) {\rm log} \left( 1 +
M \right) \right),
\end{equation}
where
$$
P_{4}(M) = \frac{1}{5} M^{4} + \frac{35}{4}M^{3} + 37 M^{2} +
\frac{99}{2} M + 21,
$$
$$
M = \frac{| 1 + u p |^{2}}{u(1-u)} = - |p|^2 +
\frac{|1+p|^{2}}{1-u} + \frac{1}{u} \ , \qquad p = \frac{1+q}{1-q} \in
{\mathbb H}.
$$
We shall not write down the explicit expression for \Ref{corrq}. We
only mention that it is a sum of rational functions of $|q|^2$,
$|p|^2$, $|1-q|^2$ multiplied by logarithms and dilogarithms of $|q|$,
$|1-q|$ etc.  The logarithms and dilogarithms of $q$, $1-q$, etc., in
\Ref{corrq} imply that the four-dimensional theory is a logarithmic
conformal field theory. For more details, see \cite{FLNii}.

\section{New observables}    \label{new obs}

In this section we explain how to extend our analysis to more
general observables, such as those corresponding to vector fields and
differential operators on the target manifold.

\subsection{The enhancement of the space of observables.} 

The observables that we have studied so far are mostly the evaluation
observables corresponding to differential forms on the target
manifold.  The novelty of our approach is to consider the pull-backs
of {\em all} differential forms, not necessarily the closed ones
(which correspond to the BPS observables comprising the topological
sector of the theory). This allowed us to see the previously hidden
logarithmic structure of the states and operators in the quantum field
theory with instantons.  Now we consider an entirely new class of
observables; namely, those corresponding to the vector fields and,
more generally, differential operators on the target manifold. These
observables are invisible in the BPS sector. We now explain how to
define these observables in the models at the special point $\tau^{*}
= -i \infty$. One motivation to study them is that the deformation of
our models back to finite values of $\tau^*$ is achieved using these
observables.

The idea is to deform the instanton equations and to study the response
of the correlation functions of evaluation observables to this
deformation.  This dependence of the correlation functions on the
deformations may then be interpreted as one caused by the insertion in the
correlation function of a new type of observables. Note that the
correlation function of closed evaluation observables is independent
of the deformations -- therefore, it is crucial to include the
evaluation observables of non-closed differential forms. It turns out
that in this way we may generate the correlation functions of all
local observables in the neighborhood of a twisted supersymmetric
point in the space of field theories. Thus, we arrive at a
perturbative definition of the path integral of our model without
violating the intrinsically non-linear structure of the space of
fields.  We explain in the examples below how this method works for
finite-dimensional integrals and in the case of quantum mechanics. A
more thorough treatment will be presented in \cite{FLNii}.

\subsection{Finite-dimensional case.} 

Path integral is usually ``defined'' by a formal extension to the
infinite-dimensional case of some procedures that are well-defined for
the integrals over finite-dimensional spaces. For example, the
well-known Feynman diagram approach to the path integral is based on
the relation
\begin{multline}
\int_{{\R}^{n}} {\rm d}^{n} \ x \ \exp \left(- {\half} \left(x, Ax
\right) + W(x)+(t,x) \right) =\\
\qquad\qquad \exp \left( W \left( {{\pa}\over {\pa}t} \right)\right)
\int_{{\R}^{n}} {\rm d}^{n} \ x\  \exp
\left(- {\half} \left(x, Ax \right)  +( t, x) \right)
\label{wab}
\end{multline}
and the well-known expression for the Gaussian integral on the right
hand side. Here we consider $W(x)$ as a polynomial with formal
coefficients, $x \in {\R}^{n}$, and $t$ belongs to the dual vector
space $({\R}^{n})^*$.

Now we propose to start with another exact relation. Let $X$ be a
finite-dimensional manifold, $V \to X$ a vector bundle over $X$, and
$v$ a section of $V$. Then this relation is
\begin{equation}
\int {\rm d}p_a {\rm d}\pi_a {\rm d}x^i {\rm d}\psi^i \
\exp \left( i p_a v^a(x) + i \pi_a \partial_j v^a  \psi^j \right)
F(x,\psi) = \int_{{\rm zeroes\ of} \ v} \omega_F
\label{euler}
\end{equation}
where $\omega_F$ denotes the differential form on $X$ corresponding to
the function $F$ on the $\Pi TX$ (with even coordinates $x^{i}$ and
odd coordinates $\psi^{i}$).  The variables $p_a$ and $\pi_a$
correspond to the even and odd coordinates on $V$.

Let us now deform $v$. In other words, let
\begin{equation}
v_{\eps} = v_{0} + {\eps}^{\alpha} v_{\alpha},
\end{equation}
where $v_{0}$ and $v_{\al}$ are section of $V$, and ${\eps}^{\al}$ are
(formal) deformation parameters. Consider the relation \Ref{euler} in
which $v$ is replaced by $v_\ep$. We will consider the right hand side
of \Ref{euler} as the {\em definition} of the generating function for
the integrals of polynomials in the new observables ${\CO}(v_{\al})$
corresponding to the $v_\al$'s. In other words, we define the
correlation function
$$
\left\langle {\CO}_{F} e^{{\eps}^{\al} {\CO}(v_{\al})}
\right\rangle
$$
involving the old evaluation observables ${\CO}_{F}$ corresponding to
differential forms and the new ones, ${\CO}(v_{\al})$, as
\begin{equation}    \label{eulerd}
\int {\rm d}p_a {\rm d}\pi_a
{\rm d}x^i {\rm d}\psi^i \ \exp \left( i p_a v^a_{\ep}(x) + i \pi_a
\partial_j v^{a}_{\ep} \psi^j \right)\ F(x,\psi) = \int_{{\rm zeroes\
    of} \ v_{\eps}} \omega_F.
\end{equation}

Now we wish to use the relation (\ref{eulerd}) as the {\em definition}
of the integral on the left hand side in the case when $X$ is
infinite-dimensional, provided that the dimension of the space of
zeroes of the section $v$ is finite (so that the right hand side of
(\ref{eulerd}) makes sense). This is exactly the situation that we
encounter in our instanton models. Indeed, the instanton models
discussed above (which appear in the limit $\tau^* \to -i \infty$) are
described by first order Lagrangians. Therefore the path integral in
these models is an infinite-dimensional version of the integral on the
left hand side of \eqref{eulerd}. The analogue of formula
\eqref{eulerd} then becomes the statement that the correlation
functions in these models localize on the finite-dimensional moduli
space of instantons, which is interpreted as the space of zeroes of an
appropriate section of a vector bundle over the space of all fields.

Once we have defined the path integral in this way, it is natural to
ask: what happens if we deform the instanton moduli space? This means
deforming the section $v_0$, which defines our moduli space, by adding
to it ${\eps}^{\alpha} v_{\alpha}$. The point is that the result
should be interpreted as the insertion of a new observable
$\on{exp}({\eps}^{\alpha} {\CO}(v_{\alpha}))$ into the path integral
(corresponding to the initial instanton moduli space).

Note that in the case when the zeroes of $v_{0}(x)$ are isolated, the
definition \Ref{eulerd} reproduces the same results as the traditional
perturbative approach. However, unlike the traditional approach,
our definition does not require that we choose any linear structure
on the space $X$.  Therefore it is well-adapted to strongly non-linear
systems such as the sigma-models and non-abelian gauge theories.

We will now illustrate how this works in a toy model example of
quantum mechanics.

\subsection{Quantum mechanical example.}

We take as $X$ the space of maps of the circle ${\bS}^{1}_{t}$ (with
the coordinate $t \sim t+1$) to another circle ${\bS}^{1}_{q}$ (with
the coordinate $q \sim q +1$).  As the vector bundle $V$ we take the
bundle whose fiber at $q(t): {\bS}^{1}_{t} \to {\bS}^{1}_{q}$ is
\begin{equation}
V_{q(t)} = {\Gamma} \left( q^{*}T{\bS}^{1}_{q} \otimes
{\Omega}^{1}({\bS}^{1}_{t}) \right).
\label{sect}
\end{equation}
As a section of this bundle we take:
\begin{equation}
v_{0} = \left( dq - 1 \cdot dt \right),
\label{sec}
\end{equation}
where $1$ is understood as a particular vector field ${\pa}_{q}$ 
on the target ${\bS}^{1}$. We will consider deformations $v_\ep = v_0
+ {\eps}^{\alpha} v_{\alpha}$, where
\begin{equation}
v_{\al}  = 1\cdot u_{\al}(t) dt,
\label{ual}
\end{equation}
with the restriction that
$$
\int_{0}^{1} u_{\al}(t) dt = 0.
$$
The space of zeroes of the vector field $v_{\eps}$ is
the space of solutions of the equation
\begin{equation}
q (t) =q_{0} + t+{\eps}^{\al} \int_0^t u_{\al}(t^{\prime})dt^{\prime}
\label{qtqo}.
\end{equation}
For the observable $F = \exp 2{\pi} i \left( q \left( t_{2} \right)-
q\left( t_{1} \right) \right) \delta\left(q\left(0\right)\right) {\psi}(0)$
the integral \Ref{eulerd} is given by the formula
\begin{multline}
\int {\mc D}p(t) {\mc D}\pi(t) {\mc D}q(t) {\mc D}\psi(t) e^{i \int (
  p \left( dq - dt \right) +\pi d\psi)} e^{- i \int \ p {\eps}^{\al}
  u_{\al} (t) dt} \cdot \\ \cdot e^{2{\pi}i ( q(t_{2})-q(t_{1}))}
  \delta(q(0)) \psi(0) = e^{2{\pi} i \left( t_{2}-t_{1}+ {\eps}^{\al}
  \int_{t_1}^{t_2} u_{\al} (t) dt \right)} \ .
\label{longf}
\end{multline}
Now we take
\begin{equation}
{\eps}^{\al}u_{\al} (t)={\eps} (\delta(t-t_{+})- \delta(t-t_{-})),
\label{epdel}
\end{equation}
where
\begin{equation}
t_+ > t_2 > t_- > t_1 \ .
\label{order}
\end{equation}
We obtain from \Ref{longf} the following correlation
function:\footnote{For the general ordering of times, not necessarily
agreeing with \Ref{order}, $e^{2\pi i \eps}$ will be replaced by
$\exp(2{\pi}i \eps \on{link}([t_+]-[t_-], [t_2]-[t_1]))$, where
$\on{link}(A,B)$ is the linking number of the $0$-chains $A$ and
$B$.}
\begin{equation} \label{linking}
\langle e^{\eps p(t_+)} e^{2{\pi}i q(t_2)} e^{-\eps p(t_-)}
e^{-2{\pi}i q(t_1)} \delta(q(0)) {\psi}(0) \rangle = e^{2{\pi} i
(t_{2}-t_{1})} e^{2{\pi}i \eps}
\end{equation}
Now, from
(\ref{linking}) it follows that
$$
e^{\eps p} e^{2{\pi}i q} e^{-\eps p} e^{-2{\pi} i q} = e^{2{\pi}i \eps} \ .
$$
Thus, we see that the above formalism reproduces the 
Heisenberg relation in quantum mechanics on a circle. In order to
reproduce the Heisenberg relation in its standard form we should
consider maps from ${\R}^{1}$ to ${\R}^{1}$ in a similar fashion.

\subsection{Generalization to two and four dimensions}

Applying a similar formalism in two- and four-dimensional instanton
models allows us to introduce new observables in these models and
opens the door for a perturbation theory away from the special point
$\tau^* = -i \infty$ towards the physical range of the coupling
constants.

For example, in two-dimensional sigma models these observables have
the form
\begin{multline}    \label{O v}
{\mc O}(v) = V^j(x^i,\ol{x}^{\ol{i}}) p_{jw} +
W^{\ol{j}}(x^i,\ol{x}^{\ol{i}}) p_{\ol{j}\ol{w}} \\ + {\psi}^{i}
(\pi_{jw} \pa_{x^i}V^j + \pi_{\ol{j}\ol{w}}) \pa_{x^i}W^{\ol{j}} +
\ol{\psi}^{\ol{i}} (\pi_{jw} \pa_{\ol{x}^{\ol{i}}}V^j +
\ol{\pi}_{\ol{j}\ol{w}} \pa_{\ol{x}^{\ol{i}}}W^{\ol{j}}).
\end{multline}
Hence they correspond to the Lie derivatives with respect to the
vector fields
$$
v = V^j \pa_{x_j} + W^{\ol{j}} \pa_{\ol{x}^{\ol{j}}}
$$
on the target manifold $X$. In particular, if this vector field is
holomorphic, then the corresponding observable belongs to the chiral
algebra of the sigma model (the chiral de Rham complex).

Why do we care about including these observables? We would like to
understand our models in the vicinity of the special point $\tau^* \to
-i \infty$. In the case of sigma models, for example, deformation to
finite values of $\tau^*$ is achieved by adding to the Lagrangian the
term (where $G^{i \ol{j}}$ is a constant matrix)
\begin{equation}
G^{i\ol{j}} {\mc O}( e_{i}) {\mc O}(e_{\ol{j}}) \ ,
\label{hee}
\end{equation}
where $e_{i}$, $e_{\ol{i}}$ are the vierbein components, $G^{i \ol{j}}
e_{i}^{\m} \otimes e_{\ol{j}}^{\nu}$ being the inverse metric on
$X$. As \Ref{hee} is bilinear in the operators \eqref{O v}, we see
that it is the observables of the form \eqref{O v} that are needed in
order to define the deformation of our models to the physical range of
coupling constants.  Implementing this program will allow us to define
the correlation functions in our quantum models entirely in terms of
finite-dimensional integrals.

\vspace*{-5mm}

\section{Remarks and outlook}

We now summarize what we have learned so far.  In quantum field
theories in one, two, and four space-time dimensions we study the
limit, in which the loop counting parameter is sent to zero, the theta
angle has acquired a large imaginary part, so that a particular
combination, the instanton action, is kept finite, while the
anti-instanton action is sent to infinity. The resulting theory is
solvable and can be used as the starting point of a perturbation
theory. The simplest models are those in which the path integral
measure is canonically defined. Such a theory is obtained from the
physical supersymmetric theory with ${\mathcal N}=2$ supersymmetry by
the procedure known as twisting. The resulting theory has a nilpotent
symmetry, generated by a scalar supercharge ${\mathcal Q}$, on any
worldsheet. However, we look at the full quantum field theory, beyond
its topological sector. Let us summarize the salient features of this
theory.

\subsection{Logarithmic structure.} Instanton corrections induce
mixing between the approximate eigenstates of the energy operator. In
the absence of anti-instantons this mixing does not change the energy
eigenvalues. It lifts, however, the degeneration of the spectrum by
making some of the approximate eigenstates into adjoint vectors of the
Hamiltonian, thereby creating Jordan blocks. In order to observe this
structure, it is crucial to consider non-BPS observables,
i.e. those non-commuting with the supercharge ${\mathcal Q}$. The
correlation functions of BPS observables are independent on the
worldsheet positions, and probe the vacua of the theory only, on which
Jordan blocks (and hence logarithms) cannot arise.

In the quantum mechanics on a K\"ahler manifold $X$ with the
holomorphic ${\C}^{\times}$-action, whose $U(1)$ part acts
isometrically, the non-diagonal part of the Hamiltonian can be related
to the so-called Grothendieck--Cousin operators. They acts on the
spaces of delta-like differential forms supported on the strata of our
manifold (the ascending and descending manifolds of the Morse
function), sending forms on a given stratum to those on the adjacent
strata of complex codimension one.

\subsection{Space of states and the fate of Hodge theory} The
supersymmetric quantum mechanical models with finite $\la$ give us a
particular realization of the Hodge algebra: a pair of odd operators,
acting on a superspace (in this case, differential forms on a
manifold), whose anti-commutator is an elliptic even operator with
discrete spectrum.  As is well-known in the classical examples of
Hodge theory, the cohomology of any of these odd operators may be
identified with the space of harmonic forms with respect to the
elliptic operators. Moreover, if the two odd operators are Hermitian
conjugate with respect to some Hermitian pairing on the forms, the
space of all forms has an orthogonal decomposition into the exact
forms for the first operator, the exact forms for the second
operators, and the harmonic forms.

As we take the limit $\la \to \infty$, together with the conjugation
by $e^{\la f}$, the Hodge algebra degenerates to
\begin{equation}
{\mathcal L}_{v} = \{ d , \ \iota_{v} \} 
\label{cartan}
\end{equation}
Now it is no longer true that the "harmonic" forms (the ground states
of the quantum mechanical system in the $\la = \infty$ limit) are
annihilated by $d$ and $\iota_{v}$. But they are annihilated by their
commutator.

If $X$ is a real manifold and $V$ is a gradient vector field of a
general Morse function $f$, then the typical ground states will be the
differential delta-forms $\Delta_\al$ (see formula
\eqref{Delal}). These are distributions (or currents) supported on the
cells of the Morse cell decomposition. They are the singular
differential forms, which are delta-forms in the directions
transversal to the corresponding cell, and are constant functions in
the directions along the cell. The application of the de Rham
differential to such a form produces a delta-form, supported at the
boundary of the cell.  Thus, the action of de Rham differential on the
space of ground states coincides with differential of the Morse
complex (note that it is identically equal to $0$ in the case of
K\"ahler manifolds, because the cells have even real dimensions). This
may be viewed as a reformulation of Witten's approach to Morse theory
\cite{W:morse}.

We go beyond the ground states and consider other delta-like
differential forms supported on the cells of the Morse
decomposition. We claim that together they span the space of states of
our model (actually, the spaces of ``in'' and ``out'' states, which
correspond to the descending and ascending cells, respectively). These
delta-like forms are interpreted as distributions on our manifold
defined by means of an Hadamard-Epstein-Glaser type
regularization. Because of the ``cutoff'' dependence of this
regularization, the action of the Hamiltonian becomes
non-diagonal. This is how the instanton effects are realized on the
space of states in our limit.

\subsection{Holomorphic factorization}

In the quantum mechanical model on K\"ahler manifolds with $U(1)$
isometry (loop spaces of interest fall into this category) the space
of ``in'' states ${\mc H}^{\on{in}}$ has the form
\begin{equation}
{\mc H}^{\on{in}} = \bigoplus_{\al \in A} {\mc H}^{\on{hol}}_\al
\otimes {\mc H}^{\on{anti-hol}}_\al,
\label{holfac}
\end{equation}
where $A$ is the set of fixed points of the $U(1)$ action. This is a
version of holomorphic factorization, which exhibits some familiar
features of two-dimensional conformal field theory, such as the
appearance of conformal blocks. Note that the decomposition
\Ref{holfac} is possible because the cells of the Morse decomposition
are isomorphic to complex vector spaces in the K\"ahler case. This
implies that the spaces of delta-like forms supported on these cells
decompose into tensor products of holomorphic and anti-holomorphic
ones. Indeed, a delta-like form supported on a cell may be written as
a polynomial differential form on the cell itself times a polynomial
in the derivatives in the transversal directions, applied to the
delta-form supported on the cell.

We observe a similar decomposition in the two-dimensional sigma
models with K\"ahler target manifolds.

\subsection{New invariants of manifolds?}

Going beyond the ground states (i.e., beyond the cohomology of the
${\mathcal Q}$-operator) may lead us to new invariants of
four-manifolds.  Let us elaborate on this point. The ordinary
Donaldson theory produces invariants of the smooth structure of a four
manifold ${\mathbf M}^{4}$ out of the topology of the moduli space of
gauge instantons. The latter can be viewed as the investigation of the
overlaps of the ground states of the four-dimensional theory. From
this perspective, going beyond the ground states is analogous, in a
sense, to working with the minimal model of the differential graded
algebra of differential forms on a manifold (as opposed to just the
cohomology of the manifold). This is reminiscent of D.~Sullivan's
approach to the reconstruction of the rational homotopy type of smooth
manifolds. We therefore hope that the study of the analogous algebras
of forms on the moduli spaces of gauge instantons will produce finer
invariants of four-manifolds.

\subsection{Non-supersymmetric theories}

The limit ${\tau}^{*} \to - i\infty$ may also be studied in the
context of non-supersymmetric theories. In this case the definition of
the path integral measure requires more work. The spaces of states are
defined using half-forms and their infinite-dimensional analogues. In
the case of the sigma model on a flag variety one finds an affine Lie
algebra at the critical level $k = - h^{\vee}$ as a chiral symmetry
algebra (hence this model is related to the geometric Langlands
correspondence). Since these manifolds are not Calabi-Yau, the
conformal symmetry is not preserved. Moreover, the analogues of the
logarithmic terms in the Hamiltonian generate a mass gap. These models
will be studied in Part III of \cite{FLNii}.

\bigskip

\noindent {\bf Acknowledgments.} This project was supported by DARPA
through its Program ``Fundamental Advances in Theoretical
Mathematics''. In addition, EF was partially supported by
the NSF grant DMS-0303529; ASL by the Russian Federal Agency of
Atomic Energy, by the grants RFBR 07-02-01161, INTAS-03-51-6346, NWO
project 047.011.2004 and NSh-8065.2006.2; and NN by European RTN
under the contract 005104 "ForcesUniverse", by ANR under the grants
ANR-06-BLAN-3$\_$137168 and ANR-05-BLAN-0029-01, and by the grants
RFFI 06-02-17382 and NSh-8065.2006.2.

The material of these notes was presented in several talks given by
the authors, in particular, at the Workshops ``Strings and Branes: The
present paradigm for gauge interactions and cosmology'' in Cargese,
``Homological Mirror Symmetry'' at ESI, Vienna, and Strings--2006 in
Beijing, in May--June of 2006. We thank the organizers of these
workshops for the invitations.

\end{document}